\documentclass[3p, twocolumn, 10pt]{elsarticle}
\usepackage[T1]{fontenc}
\usepackage[english]{babel}

\usepackage{hyperref}

\usepackage{amsmath}
\usepackage{amsfonts,amssymb}

\usepackage[T1]{fontenc}
\usepackage{bm}

\usepackage{graphicx}
\usepackage{lscape}
\usepackage{subfig}
\usepackage{float,caption}
\usepackage{mathtools}

\newcommand*{\GtrSim}{\smallrel\gtrsim}
\newcommand*{\LessSim}{\smallrel\lesssim}

\makeatletter
\def\ps@pprintTitle{%
  \let\@oddhead\@empty
  \let\@evenhead\@empty
  \def\@oddfoot{\reset@font\hfil\thepage\hfil}
  \let\@evenfoot\@oddfoot
}
\newcommand*{\smallrel}[2][.8]{%
  \mathrel{\mathpalette{\smallrel@{#1}}{#2}}%
}
\newcommand*{\smallrel@}[3]{%
  \sbox0{$#2\vcenter{}$}%
  \dimen@=\ht0 %
  \raise\dimen@\hbox{%
    \scalebox{#1}{%
      \raise-\dimen@\hbox{$#2#3\m@th$}%
    }%
  }%
}
\makeatother

\def\({\left(}
\def\){\right)}

\usepackage{xcolor}
\xdefinecolor{redviolet}{RGB}{228,38,207}
\xdefinecolor{brickred}{RGB}{220,60,66}
\xdefinecolor{bluscuro}{RGB}{51,51,179}

\usepackage{amssymb}

\begin{document}

\begin{frontmatter}
\title{\textbf{Single \texorpdfstring{$\pi^0$}{p0} production in \texorpdfstring{$\mu e$}{me} scattering at MUonE}}

\author[a,b]{Ettore Budassi}
\ead{ettore.budassi01@universitadipavia.it}

\author[b]{Carlo M. Carloni Calame}
\ead{carlo.carloni.calame@pv.infn.it}

\author[a,b]{Clara Lavinia Del Pio}
\ead{claralavinia.delpio01@universitadipavia.it}

\author[b]{Fulvio Piccinini}
\ead{fulvio.piccinini@pv.infn.it}

\affiliation[a]{organization={Dipartimento di Fisica, \unexpanded{Università} di Pavia},
            addressline={Via A. Bassi 6},
            postcode={27100},
            city={Pavia},
            country={Italy}}
\affiliation[b]{organization={INFN, Sezione di Pavia},
            addressline={Via A. Bassi 6},
            postcode={27100},
            city={Pavia},
            country={Italy}}

\begin{abstract}
The recently proposed MUonE experiment at CERN aims at
  providing a novel determination of the leading order hadronic
  contribution to the muon anomalous magnetic moment through the study
  of elastic muon-electron scattering at relatively small momentum
  transfer. The anticipated accuracy of the order of 10ppm demands for 
  high-precision prediction in radiative corrections to the $\mu e $ scattering as well as for robust quantitative estimates of all possible background processes. In this letter, the contribution due to the emission of a neutral pion through the process $\mu e \to \mu e \pi^0$ is studied and its numerical impact is discussed in different phase space configurations by means of the upgraded Monte Carlo event generator \textsc{MESMER}. In fact, single $\pi^0$ production could be a source 
  of reducible background for the measurement of the QED running coupling constant 
  at MUonE and it could be also an important background for possible New Physics searches at MUonE involving $2 \to 3$ processes, in phase space regions complementary to the ones characteristic of the elastic $\mu e$ scattering.
\end{abstract}

\end{frontmatter}

\section{Introduction}
\label{sec:introduction}
The value of the anomalous magnetic moment of the muon, $a_\mu = (g-2)_\mu / 2$, is a fundamental quantity in particle physics. Very recently, the measurement performed by the Fermilab Muon $g-2$ Experiment (E969)~\cite{Abi:2021gix} has been combined with the Brookhaven National Laboratory result of 2001~\cite{Bennett:2006fi}, yielding a deviation of 4.2$\sigma$ from the theoretical prediction of the muon anomaly, which stems from the QED, weak and strong sectors of the Standard Model (SM)~\cite{Jegerlehner:2009ry,Jegerlehner:2017gek}. The current status of the SM theoretical prediction for the muon $g-2$ has been recently reviewed in Ref.~\cite{Aoyama:2020ynm}. The comparison between theory and experiment provides a stringent test of the SM: indeed, a deviation from the SM expectation can point to possible New Physics signals. 

The dominant contributions to the theoretical error comes from Leading Order Hadronic Vacuum Polarisation (HLO) and Hadronic Light by Light (HLbL) effects, the latter having less impact \cite{Aoyama:2020ynm}.
Concerning the HLO term, the BMW collaboration has recently presented a precise determination of 
$a_\mu^\text{HLO}$, based on Lattice QCD calculations, with an uncertainty of
$0.78\%$~\cite{Borsanyi:2020mff} and a central value larger than the ones
obtained via dispersive techniques, which typically allow to achieve the highest available precision in $a_\mu^\text{HLO}$. 
The discrepancy between the BMW $a_\mu^\text{HLO}$ estimate, which would lead to an $a_\mu$ prediction in agreement with the BNL and FNAL experimental determinations~\cite{Abi:2021gix,Bennett:2006fi}, and the $a_\mu^\text{HLO}$ results based on dispersion relations ranges from $2$ to $2.5\sigma$, depending on the reference value used for the
dispersive approach. In order to clarify this tension 
and in view of the importance of a very robust theoretical SM prediction for $a_\mu$, alternative and independent methods for the evaluation of $a_\mu^\text{HLO}$ are therefore more than welcome, if not necessary.

Recently, a novel approach has been proposed in Ref.~\cite{Calame:2015fva}
to derive $a_\mu^\text{HLO}$ from a measurement of the effective
electromagnetic coupling constant in the space-like region via
scattering data, making use of a relation between
$\Delta\alpha_\text{had}(q^2)$ at negative squared momenta and
$a_\mu^\text{HLO}$ (see also~\cite{Lautrup:1971jf})~\footnote{Very recently, analytic expressions have been provided in Ref.~\cite{Balzani:2021del} to extend the space-like 
calculation of the hadronic vacuum polarization contribution to $a_\mu$ up to next-to-leading order precision.}.
Shortly afterwards, the elastic scattering of muons on electrons has been
identified as an ideal process for such a
measurement~\cite{Abbiendi:2016xup}~\footnote{A method to measure the running of the QED coupling constant in the space-like region using small-angle Bhabha scattering was proposed in Ref.~\cite{Arbuzov:2004wp} and applied to LEP data by the OPAL Collaboration~\cite{OPAL:2005xqs}. In the time-like region, the effective QED coupling constant in the region below~1~GeV has been recently measured by the KLOE Collaboration~\cite{KLOE-2:2016mgi}.}
and a new experiment, MUonE, has been proposed at CERN to measure the
differential cross section of this process \cite{MUonE:LoI}.
In order for this new determination of $a_\mu^\text{HLO}$ to be
competitive with the traditional dispersive approach, the uncertainty
in the measurement of the $\mu e$ differential cross section must be
of the order of 10ppm, as described in Refs.~\cite{Abbiendi:2016xup,MUonE:LoI}.

Such a target precision is extremely challenging on both the experimental as well as the theoretical side. Recent progress, after the experimental proposal~\cite{MUonE:LoI}, in  the study of the main systematics and in the development of the experimental apparatus and data analysis strategies have been documented in Refs.~\cite{Abbiendi:2019qtw,Ballerini:2019zkk,Abbiendi:2021xsh,Abbiendi:2020sxw,Abbiendi:2022liz,Abbiendi:2022oks}. On the theory side several groups have already reached important milestones related to next-to-next-to-leading 
QED radiative corrections and their implementation in independent MC  codes~\cite{Mastrolia:2017pfy,Alacevich:2018vez,DiVita:2018nnh,Engel:2019nfw,Banerjee:2020tdt,CarloniCalame:2020yoz,Banerjee:2020rww,Budassi:2021twh,Bonciani:2021okt}.

In this paper we focus on hadronic corrections, whose virtual part has already been studied in Refs.~\cite{Fael:2018dmz,Fael:2019nsf}. The real-emission contributions consist of three channels: $\mu^\pm e \to \mu^\pm e \pi^+ \pi^-$,  $\mu^\pm e \to \mu^\pm e \pi^0 \pi^0$ and $\mu^\pm e \to \mu^\pm e \pi^0$~\footnote{The need for a quantitative estimate of these contributions was firstly stressed in Ref.~\cite{Banerjee:2020tdt}.}. While pion-pair production is extremely constrained 
for the limited available phase space at MuonE, since, as already noted in Ref.~\cite{Budassi:2021twh}, a realistic event selection makes the cross section vanishing, the single pion production channel deserves a more detailed investigation. In fact, while on dimensional grounds the impact of pion emission in $\mu e$ scattering is expected to be suppressed w.r.t. the elastic scattering, 
the $\pi^0$ production is dynamically enhanced in the region of small electron and muon scattering angles. In this region, which is  particularly interesting for MUonE, the elastic cross section becomes negligible and the sensitivity to the running of the electromagnetic coupling constant reaches its maximum. Since 
the MUonE experimental apparatus relies primarily on 
high-resolution charged track reconstruction, the two photons which the $\pi^0$ decays into are not detected. Hence, the signature of the process under study corresponds to the one of  the elastic $\mu^\pm e \to \mu^\pm e$ process: this fact reflects the importance of studying the $\pi^0$ production process.

An additional reason for a detailed study of the process $\mu^\pm e \to \mu^\pm e \pi^0$ at MUonE is its potential role as a background to possible New Physics searches in phase space regions that are complementary to the ones which characterise the elastic 
scattering. On the one hand, according to the findings of  Refs.~\cite{Masiero:2020vxk,Dev:2020drf}, $\mu e$ elastic scattering at MUonE is insensitive to possible New Physics contamination. On the other hand, at MUonE it could be possible to have sensitivity to the production of a light new gauge boson $Z^\prime$ through the 
process $\mu^\pm e \to \mu^\pm e Z^\prime$~\cite{Asai:2021wzx} or 
to the production of a dark photon through the process 
$\mu^\pm e \to \mu^\pm e A^\prime$~\cite{Galon:2022xcl}. 
It has to be noted that for this kind of searches additional future studies on the experimental apparatus may be probably needed. 

In Section~\ref{sec:calculation} we give a brief description of the tree-level matrix element and phase space (PS) calculation, which are implemented as an additional channel in the Monte Carlo generator MESMER~\cite{Alacevich:2018vez,CarloniCalame:2020yoz,Budassi:2021twh}~\footnote{An up-to-date version of \textsc{MESMER} can be found at \url{www.github.com/cm-cc/mesmer}.} and in Section~\ref{sec:numres} we present some numerical results for the two above mentioned complementary event selections: in Section~\ref{subsec:muoneback} 
the differential distributions relevant for the background to the 
$\mu e$ elastic scattering are examined, while Section~\ref{subsec:zpback} presents few results relevant for the event selection envisaged in Ref.~\cite{Asai:2021wzx} for the search of a light $L_\mu-L_\tau$ gauge boson. 
Section~\ref{sec:conclusion} summarises 
our study. 

\section{Calculation}\label{sec:calculation}

The process under study is: 
\begin{equation}\label{process}
    \mu^\pm(p_1) e (p_2) \to \mu^\pm(p_3) e (p_4)+ \pi^0 (p_5),
\end{equation}
where $p_i$ are the four-momenta of the particles.

\begin{figure}[h!]
    \centering
    \includegraphics[width=0.85\columnwidth]{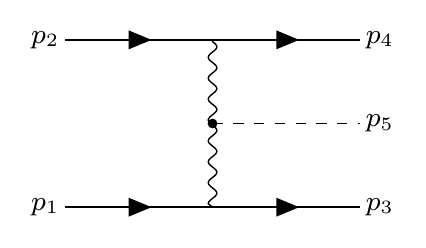}
    \caption{The tree-level Feynman diagram for the process $\mu e \to \mu e \pi^0$.
    The particle labels follow the convention of Eq. \ref{process}.
    }
    \label{fig:feynd}
\end{figure}

In order to determine the numerical impact of the $\mu^\pm e \to \mu^\pm e \pi^0$ process, the tree-level scattering amplitude and the three-body PS have been  calculated and implemented in  the Monte Carlo event generator \textsc{MESMER}. To derive the former, we started from the $\pi^0\gamma\gamma$ interaction Lagrangian 
density:
\begin{equation}
\label{eq:lagrangian}
    \mathcal{L}_{\rm{I}}=\frac{g}{2!}\varepsilon^{\mu\nu \kappa\lambda}F_{\mu\nu}F_{\kappa\lambda}\varphi_\pi ,
\end{equation}
where a form factor $F_{\pi^0\gamma^*\gamma^*}(p^2,q^2)$ dependent  
on the photon virtualities, $p^2$ and $q^2$, is understood,  
as in Ref. \cite{Brodsky:1971ud}.
We then computed the vertex Feynman rule,  $-4 i g\varepsilon_{\mu\rho\nu\sigma}p^\rho q^\sigma F_{\pi^0\gamma^*\gamma^*}(p^2,q^2)$,  where $p^\rho$ and $q^\sigma$ are the four-momenta of the photons. To obtain the $\pi^0\gamma^*\gamma^*$ coupling value, 
we exploited the relation between $g$ and the $\pi^0$ decay 
width 
\begin{equation}
\label{eq:coupling}
    g^2=\frac{4\pi\Gamma_{\pi^0\to\gamma\gamma}}{m_{\pi^0}^3} ,
\end{equation}
as in Ref. \cite{Brodsky:1971ud}. In Eq.~(\ref{eq:coupling}) $m_{\pi^0}$ is the mass of the neutral pion, $134.9766$~MeV~\cite{ParticleDataGroup:2016lqr},  
and the decay width is related to the $f_\pi$ parameter by: \[\Gamma_{\pi^0\to\gamma\gamma} = \frac{\alpha^2 m_{\pi^0}^3}{64\pi^3 f_\pi^2}.\]
By using for $f_\pi$ the value adopted in fit 1 in Table II of Ref.~\cite{Czyz:2017veo}, $f_\pi=0.092388$ GeV, the corresponding $\pi^0$ width is $\Gamma_{\pi^0\to\gamma\gamma}=7.731$ eV, to be compared with the experimentally measured value $\Gamma_{\pi^0\to\gamma\gamma}=
7.802\pm0.052\pm0.105$~eV~\cite{ParticleDataGroup:2018ovx}. 
The exact tree-level matrix element for the process $\mu e \to \mu e \pi^0$  
(Fig.~\ref{fig:feynd} shows the corresponding Feynman diagram) has been obtained by means of the symbolic manipulation 
program~\textsc{FORM}~\cite{Vermaseren:2000nd,Kuipers:2012rf,Ruijl:2017dtg}, keeping all finite mass contributions.

From now on we do not specify the electric charge of the incoming muons because 
the cross section with unpolarized muons is the same 
for $\mu^+$ and $\mu^-$. 

The form factor $F_{\pi^0 \gamma^* \gamma^*}$ has been 
calculated according to the resonance chiral symmetric model with $SU(3)$ breaking of Ref.~\cite{Czyz:2017veo}. In particular, we implemented the expression of Eq.~(14) of 
Ref.~\cite{Czyz:2017veo} with three octets and the values 
of the parameters taken from Table II, column fit 1, of the 
same reference. 

Concerning the PS parameterisation, we decomposed the three-body 
Lorentz-invariant phase space 
\begin{equation}\label{dphin}
d\Phi_3^{\text{\textsc{lips}}}=\int\prod_{i=3}^5\frac{d^3 p_i}{(2\pi)^3 2E_i}\delta^4\left(P-\sum_{j=3}^5 p_j\right).
\end{equation}
according to the following chain:
\begin{multline}
d\Phi_3^{\text{\textsc{lips}}}=(2\pi)^3\int dQ^2 d\Phi_2(P\to p_3+Q)\\
\times d\Phi_2 (Q\to p_4+p_5),    
\end{multline}
where the particle labels $3$, $4$ and $5$ stand for final state 
$\mu$, $e$ and $\pi^0$, respectively. 
The set of independent variables generated for this parameterisation is given by:
\begin{equation*}
    \vartheta_\mu,\, \varphi_\mu, \,Q^2, \,\vartheta_e^*, \,\varphi_e^* ,
\end{equation*}
where $\vartheta_e^*$ and $\varphi_e^*$ are generated in the rest frame of the pair 45 ($\vec{p}_4+\vec{p}_5=\vec{0}$).
The sampling of the variables $\vartheta_\mu$ and $\vartheta_e^*$ follows the same approach of Ref.~\cite{Budassi:2021twh}, inspired by Refs.~\cite{Berends:1994pv,Kersevan:2004yh}.

\begin{figure*}
    \centering
    \includegraphics[width=\textwidth]{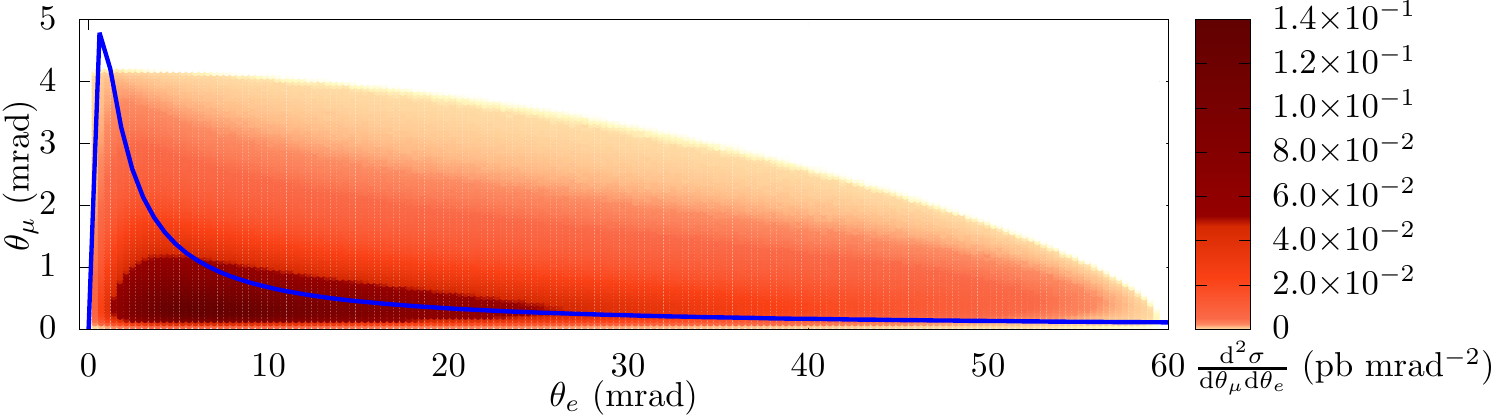}
    \caption{The configuration space for the $\mu e \to \mu e \pi^0$ process in terms of the outgoing particle angles $\vartheta_e$ (on the $x$ axis) and $\vartheta_\mu$ (on the $y$ axis) with \emph{basic acceptance cuts}. The colour gradient represents the doubly differential cross section $d^2 \sigma/d\vartheta_e d\vartheta_\mu$ (in pb mrad$^{-2}$): the darker the colour, the higher the cross section. In blue, we show the elasticity curve as defined in Eq. (5.1) of Ref.~\cite{Budassi:2021twh}.}
    \label{fig:scatterplot}
\end{figure*}

As a cross check of our Monte Carlo results, we reproduced the results of Table~1 of Ref.~\cite{Brodsky:1971ud} on the total cross section of the process $e^+ e^- \to e^+ e^- \pi^0$, by setting 
$m_\mu = m_e$ and adjusting the incoming $e^+$ beam energy to reproduce the centre of mass energies of Table 1 of Ref.~\cite{Brodsky:1971ud}. It is worth mentioning that the state of the art for the Monte Carlo simulation 
of the process $e^+ e^- \to e^+ e^- \pi^0$ is represented by the 
event generator \textsc{EKHARA}~\cite{Czyz:2010sp,Czyz:2018jpp}. 
By means of a tuned comparison, we obtained excellent agreement between 
our prediction for the ``$t$-channel'' cross section of the process $e^+ e^- \to e^+ e^- \pi^0$ and the results for the same cross section obtained with \textsc{EKHARA}. 

The above described calculation has been implemented in \textsc{MESMER} and in an independent Monte Carlo program for the cross-check of the numerical results, which we present in the following section. 

\section{Numerical Results}\label{sec:numres}

In this section we investigate the numerical impact of $\pi^0$ production in $\mu e$ scattering for typical running conditions and event selections of the MUonE experiment. 
In particular, we consider in Section~\ref{subsec:muoneback} the process 
$\mu e \to \mu e \pi^0$ as a possible reducible 
background to the determination of the running of the electromagnetic coupling constant  in the space-like channel through the measurement of the differential cross section of the elastic process $\mu e \to \mu e$. In Section~\ref{subsec:zpback} 
we discuss the impact of $\pi^0$ production as a possible background to New Physics 
searches in $2 \to 3$ channels.

\subsection{Background to \texorpdfstring{$\mu e$}{me} 
scattering}\label{subsec:muoneback}

The numerical impact of $\pi^0$ production in $\mu e$  scattering at MUonE is expected to be suppressed with respect to the $\mu e$ elastic scattering: indeed, by dimensional  analysis, the pion production cross section is reduced w.r.t. the tree level one at least by a factor of the 
order of $g^2 m_\pi^2 \sim 10^{-6}$, where $m_\pi$ is the $\pi^0$ mass and $g$ is the $\pi^0 \gamma^* \gamma^*$ coupling. 

Using the same numerical values for the parameters $\alpha$, $m_e$ and $m_\mu$ as in 
Refs.~\cite{Alacevich:2018vez,CarloniCalame:2020yoz,Budassi:2021twh} and 
$m_\pi=134.9766$~MeV~\cite{ParticleDataGroup:2016lqr}, 
the total cross section for the process $\mu e \to \mu e\pi^0$ with incoming muon energy of $150$~GeV and initial-state electron at rest is:
\[\sigma_{\mu e \pi^0}= 6.53589(6)\,\text{pb}.\]
With the two different event  selections considered in previous studies on 
radiative corrections to the elastic  process~\cite{Alacevich:2018vez,CarloniCalame:2020yoz,Budassi:2021twh}, namely 
\begin{itemize}
    \item \emph{basic acceptance cuts}: $\vartheta_\mu\LessSim 4.84$~mrad, $E_\mu\GtrSim10.28$ GeV, $\vartheta_e<100$~mrad and $E_e>0.2$~GeV,  
    \item the same basic acceptance cuts, but with $E_e>1$~GeV,
\end{itemize}
we obtain 
\[\sigma_{\mu e \pi^0}^{\text{0.2 GeV}}= 2.69836(4)\,\text{pb},\]
and
\[\sigma_{\mu e \pi^0}^{\text{1 GeV}}= 1.61597(3)\,\text{pb}, \]
respectively. 
With a tree-level elastic cross section $\sigma \sim 1265$~$\mu$b for $E_e > 0.2$~GeV 
and $\sigma \sim 245$~$\mu$b for $E_e > 1$~GeV, it is unlikely that $\pi^0$ production is numerically 
relevant for the determination of the running QED coupling constant at MUonE. 
However, in order to completely exclude possible enhancements of $\pi^0$ production in some phase space regions~\footnote{For instance, in the region $\vartheta_e \to 0$, where the elastic cross section becomes vanishingly small~\cite{Alacevich:2018vez}.}, we present also numerical results at the differential level.  
In Fig.~\ref{fig:scatterplot} we present the doubly differential distribution 
$d^2\sigma / d\vartheta_e d\vartheta_\mu$, where, by inspection, the leading 
contribution to the cross section is concentrated in the region 
$\vartheta_\mu \LessSim 1$~mrad and $\vartheta_e \LessSim 25$~mrad, which overlaps 
with the correlation curve $\vartheta_\mu(\vartheta_e)$ of the $\mu e$ elastic scattering process (blue line) for electron scattering angles larger than 
about $5$~mrad. 

\begin{figure}[t]
    \centering
    \includegraphics[width=\columnwidth]{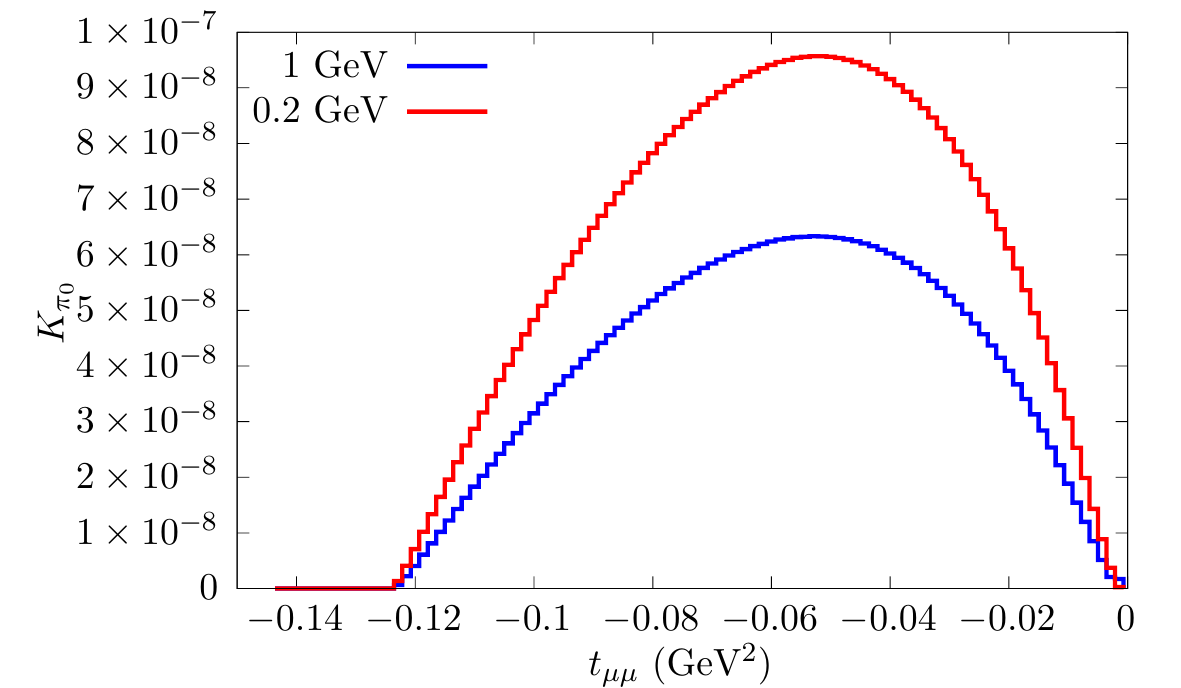}
    \caption{The \emph{K factor}, $K_{\pi^0}$, as defined in Eq.~(\ref{eq:kpi0def}) plotted against the transferred momentum along the muon line $t_{\mu\mu}$.}
    \label{fig:t13}
\end{figure}
\begin{figure}[b]
    \centering
    \includegraphics[width=\columnwidth]{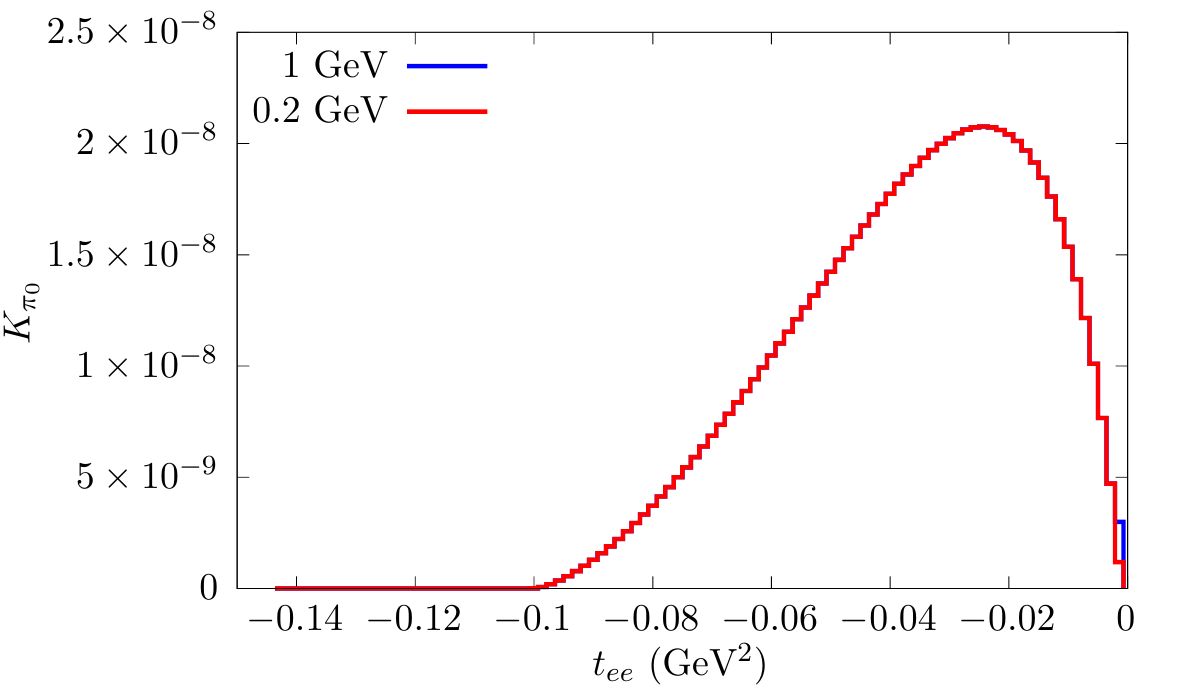}
    \caption{The same as in Fig.~\ref{fig:t13} but plotted against the squared momentum transfer along the electron line $t_{ee}$.}
    \label{fig:t24}
\end{figure}

In order to clearly illustrate the contribution of $\pi^0$ production w.r.t. the elastic $\mu e$ scattering for some key differential distributions, {\emph i.e.} 
electron and muon scattering angles in the laboratory frame and squared transferred 
momentum along the electron line $(t_{ee})$ and along the muon line $(t_{\mu \mu})$,  we consider the differential ratio of the $\pi^0$ production cross section with the  tree-level prediction for $\mu e \to \mu e$:
\begin{equation}
    K_{\pi^0}=\frac{d \sigma_{\mu e \pi^0}}{d \sigma_{\mu e}}
    \label{eq:kpi0def}.
\end{equation}

In all the remaining figures presented in this section, the blue 
histograms display $K_{\pi^0}$ calculated with the threshold 
on the electron energy of $1$~GeV while the red histograms refer to the electron energy threshold of $0.2$~GeV.

Fig.~\ref{fig:t13} shows $K_{\pi^0}$ plotted against $t_{\mu\mu}$, where the $\pi^0$ production weighs at most about $10^{-7}$ and $6\times10^{-8}$ with $E_e>0.2$ GeV and $E_e>1$ GeV, respectively. The effect remains well below 10 ppm also in the \emph{K factor} plotted against $t_{ee}$, as displayed in Fig.~\ref{fig:t24}. $K_{\pi^0}$ remains below $2.5\times 10^{-8}$ in the whole range of $t_{ee}$.

\begin{figure}[t]
    \centering
    \includegraphics[width=\columnwidth]{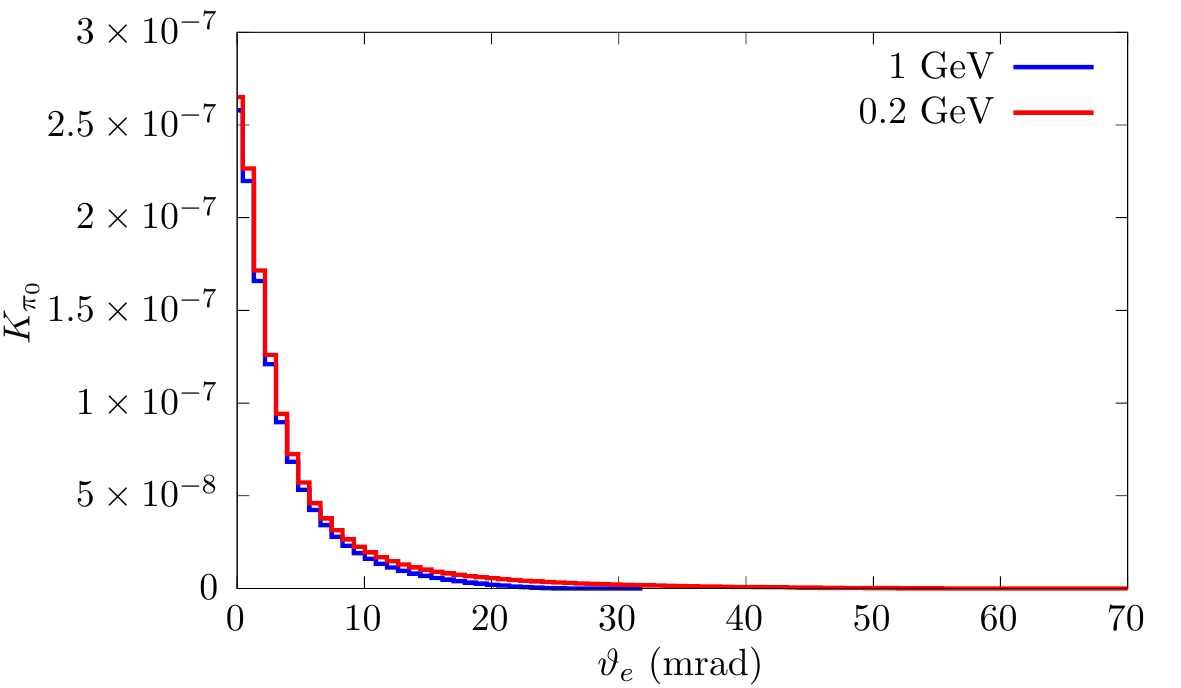}
    \caption{The same as in Fig.~\ref{fig:t13} but plotted against the outgoing electron angle in the laboratory reference frame $\vartheta_{e}$.}
    \label{fig:thetae_lab}
\end{figure}

\begin{figure}[t]
    \centering
    \includegraphics[width=\columnwidth]{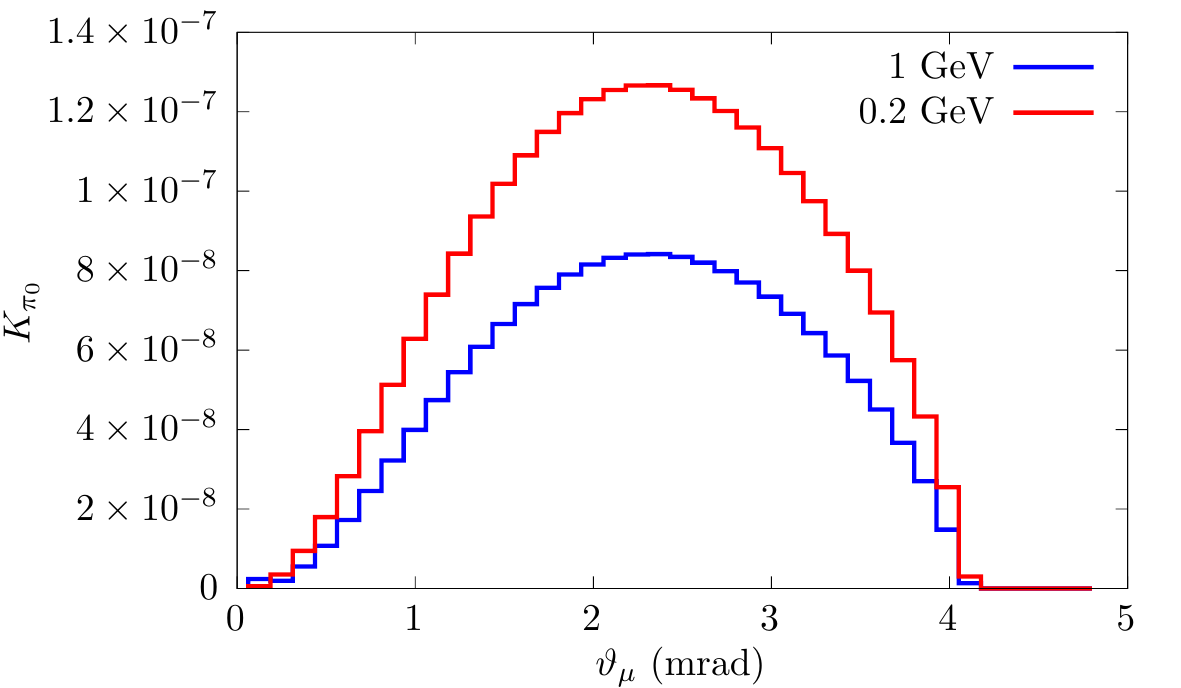}
    \caption{The same as in Fig.~\ref{fig:t13} but plotted against the outgoing muon angle in the laboratory reference frame $\vartheta_{\mu}$.}
    \label{fig:thetamu_lab}
\end{figure}

In Figs.~\ref{fig:thetae_lab} and \ref{fig:thetamu_lab}, the \emph{K factors} $K_{\pi^0}$ are shown in terms of the outgoing electron and muon angles, respectively. The scattering angles are defined in the laboratory reference frame. Also in these plots the effects of $\pi^0$ production stay below $2.6\times 10^{-7}$ and $1.3\times 10^{-7}$ respectively, and thus are clearly negligible with respect to the 10 ppm level target precision of MUonE. 

From this analysis, we can safely conclude that the process $\mu e \to \mu e \pi^0$ does not represent a  significant source of background for the extraction of the 
hadronic contribution to the running of the electromagnetic 
coupling constant at the MUonE experiment at the target level of precision.

\subsection{Background to New Physics search with \texorpdfstring{$2 \rightarrow 3$}{2->3} processes}\label{subsec:zpback}

As discussed in the Introduction, the process $\mu e \to \mu e \pi^0$ could represent a source of background for possible 
New Physics searches at MUonE in $2 \to 3$ processes, 
recently discussed, for instance, in  Refs.~\cite{Asai:2021wzx,Galon:2022xcl}. We consider 
in some detail the example suggested in Ref.~\cite{Asai:2021wzx}, within the context of a $L_\mu - L_\tau$ gauge model, where a light massive  $Z^\prime$ could be directly produced through the process $\mu e \to \mu e Z^\prime$, with subsequent decay of $Z^\prime$ to a neutrino pair. In this case, $\pi^0$ production would be a reducible background, since the pion decays into two 
photons with Branching Ratio (BR) 
$~\sim 98.8\%$~\cite{ParticleDataGroup:2018ovx}~\footnote{The $\pi^0$ production can 
be a source of reducible background to dark photon searches, as proposed in Ref.~\cite{Galon:2022xcl}, through the decay channel $\pi^0 \to e^+ e^- \gamma$.}. 
We will assume in our simulation $BR(\pi^0 \to \gamma \gamma) = 1$.

For the sake of illustration, we study the numerical impact of 
$\pi^0$ production at MUonE with the event selection suggested 
in Ref.~\cite{Asai:2021wzx}:
\begin{itemize}
    \item $\vartheta_\mu>1.5$ mrad;
    \item $E_e \in [1,25]$ GeV\, . 
\end{itemize}

This cut choice covers a PS region that is complementary to the one of interest for the 
$\mu e \to \mu e$ elastic process, discussed in the previous subsection, and could be relevant for the search of a light $Z^\prime$, 
as discussed in Ref.~\cite{Asai:2021wzx}. 
The integrated cross section for $\pi^0$ production with the above 
event selection is:
\begin{equation}
    \sigma_{\mu e \pi^0}=  0.19210(1)\, \text{pb}.
\end{equation}

Considering an integrated luminosity for the MUonE experiment of about $15$ fb$^{-1}$ as reported in Ref.~\cite{Abbiendi:2019qtw}, the number of expected events for the process $\mu e \to \mu e \pi^0$ is about $3 \times 10^3$, 
which is of the same order of magnitude of the number of signal $Z^\prime$ events reported in Ref.~\cite{Asai:2021wzx}. Of course the impact of 
$\pi^0$ production can be reduced through a photon veto strategy.
As already noted in Ref.~\cite{Asai:2021wzx}, the main reducible background is expected to come from the radiative process  $\mu e \to \mu e \gamma (\gamma)$. 
This could be studied in detail with the MC generator \textsc{MESMER}. In general, an efficient performance on photon identification and rejection would likely require some detector development with respect 
to its original proposal~\cite{MUonE:LoI}. 
\begin{figure}[h!]
    \centering
    \includegraphics[width=\columnwidth]{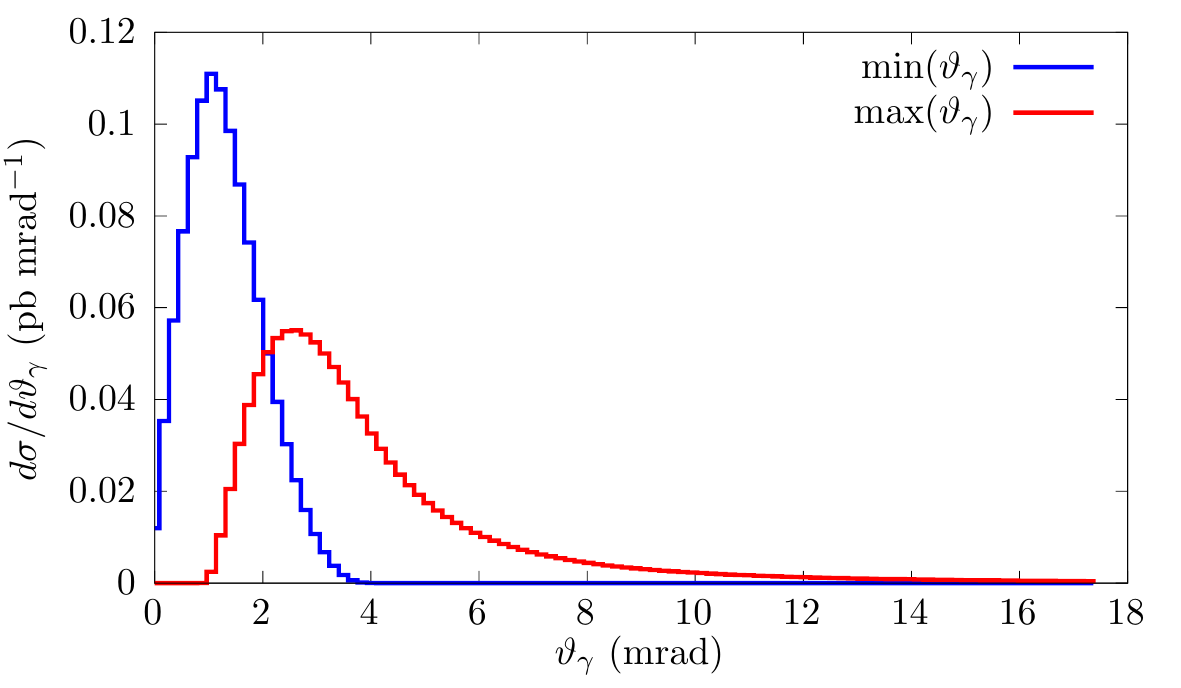}
    \caption{In blue, the differential cross section  for the $\mu e \to \mu e +\pi^{0} \to \mu e +\gamma \gamma$ process versus the minimum photon angle in the laboratory frame $\min(\vartheta_\gamma)$ is plotted. The red curve represents the same quantity but w.r.t. the maximum photon angle $\max(\vartheta_\gamma)$.}
    \label{fig:th_g1g2}
\end{figure}
In the following, for the sake of illustration, we present 
the features of the differential distributions of the photons originated from the $\pi^0$ decay, as obtained with the 
MC generator \textsc{MESMER}. In particular, we examine the differential distributions on angle and energy of the final state photons, on the invariant mass of the lepton-photon system as well as on the outgoing electron energy:

\begin{equation}
        \frac{d\sigma}{d m_{e\gamma}},\quad \frac{d \sigma}{d m_{\mu\gamma}},\quad \frac{d \sigma}{d E_{\gamma}},\quad \frac{d \sigma}{d \vartheta_{\gamma}},\quad \frac{d \sigma}{d E_e}.
\end{equation}

\begin{figure}[t]
    \centering
    \includegraphics[width=\columnwidth]{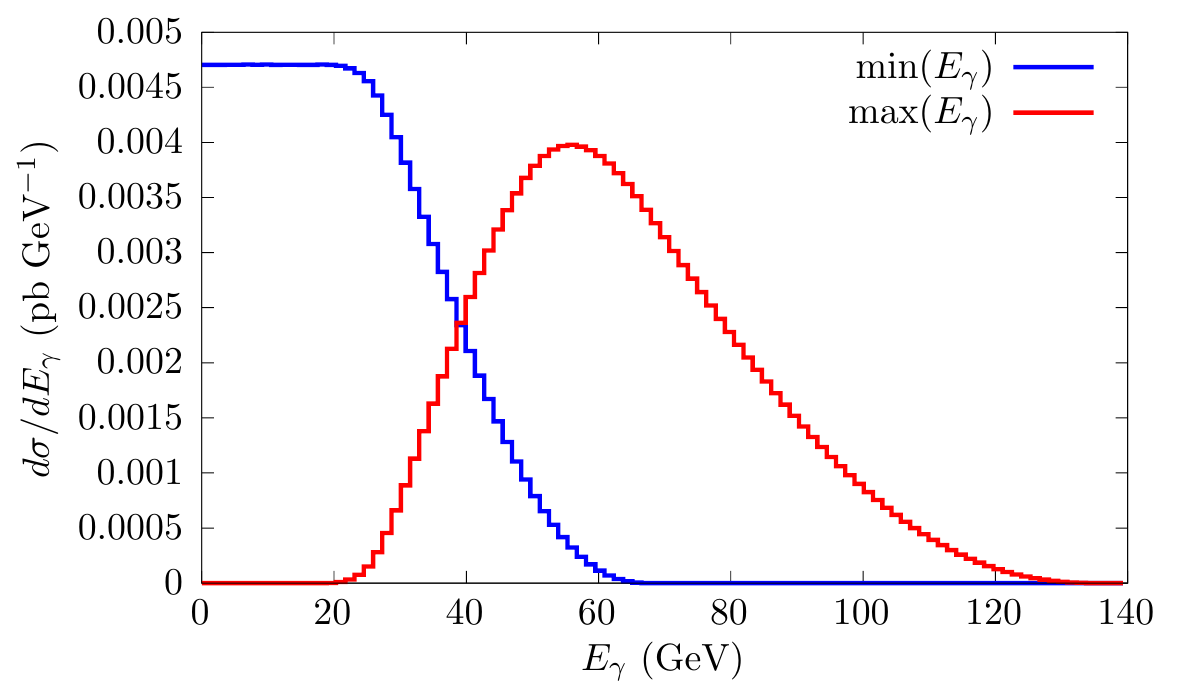}
    \caption{The same as in Fig.~\ref{fig:th_g1g2} but plotted against the minimum (blue curve) and the maximum (red curve) of the photon energy $E_\gamma$.}
    \label{fig:en_g1g2}
\end{figure}

In Fig.~\ref{fig:th_g1g2} we show the differential cross section plotted against the photon angle $\vartheta_\gamma$: it is clear that the photons are all produced in the forward region, below $\sim 10$~mrad in the laboratory reference frame. The minimum angle distribution has a very pronounced peak at about $1$ mrad where the distribution reaches approximately $0.11$ pb mrad$^{-1}$. 
Fig.~\ref{fig:en_g1g2} shows the minimum and maximum photon energy $E_\gamma$ in the laboratory frame: the former remains constant at about $4.5\times10^{-3}$ pb GeV$^{-1}$ and then goes to zero approximately at $65$ GeV. 
The maximum photon energy has about the same order of magnitude between $20$ GeV and $130$ GeV with a peak at $60$ GeV.

\begin{figure}[t]
    \centering
    \includegraphics[width=\columnwidth]{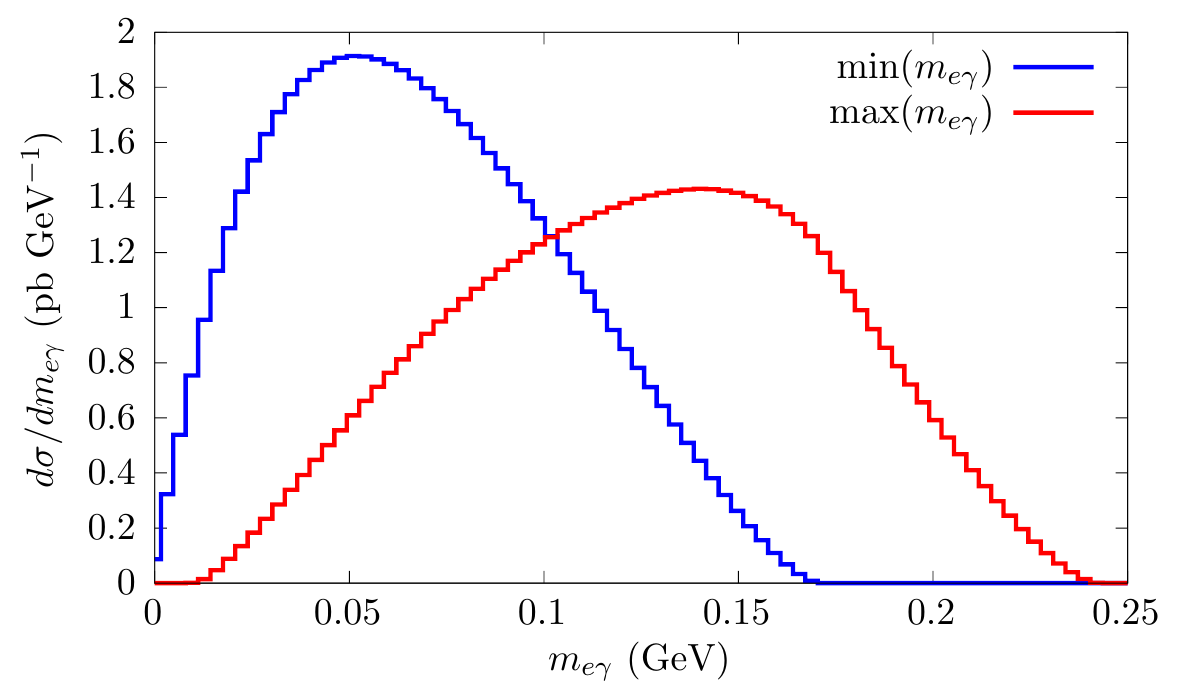}
    \caption{The same as in Fig.~\ref{fig:th_g1g2} but plotted against the minimum (blue curve) and the maximum (red curve) of the invariant mass of the $e-\gamma$ system $m_{e\gamma}$.}
    \label{fig:m_eg}
\end{figure}

We then turn our attention to the invariant mass of the lepton-photon system $m_{e \gamma}$. In Fig.~\ref{fig:m_eg} we plot the differential cross section over the minimum and maximum $m_{e \gamma}$ value: for each photon, it is calculated the invariant mass of the $e-\gamma$ system and then the minimum (maximum) is taken. The plot for the minimum value shows that $m_{e\gamma}$ stays between $0$ GeV and $0.17$ GeV with a peak that reaches about $2$ pb GeV$^{-1}$. The distribution for the maximum of the $e-\gamma$ invariant mass is more spread out, ranging from about $0$ GeV to $0.25$ GeV with a peak at $0.15$ GeV of $1.4$ pb GeV$^{-1}$.

\begin{figure}[t]
    \centering
    \includegraphics[width=\columnwidth]{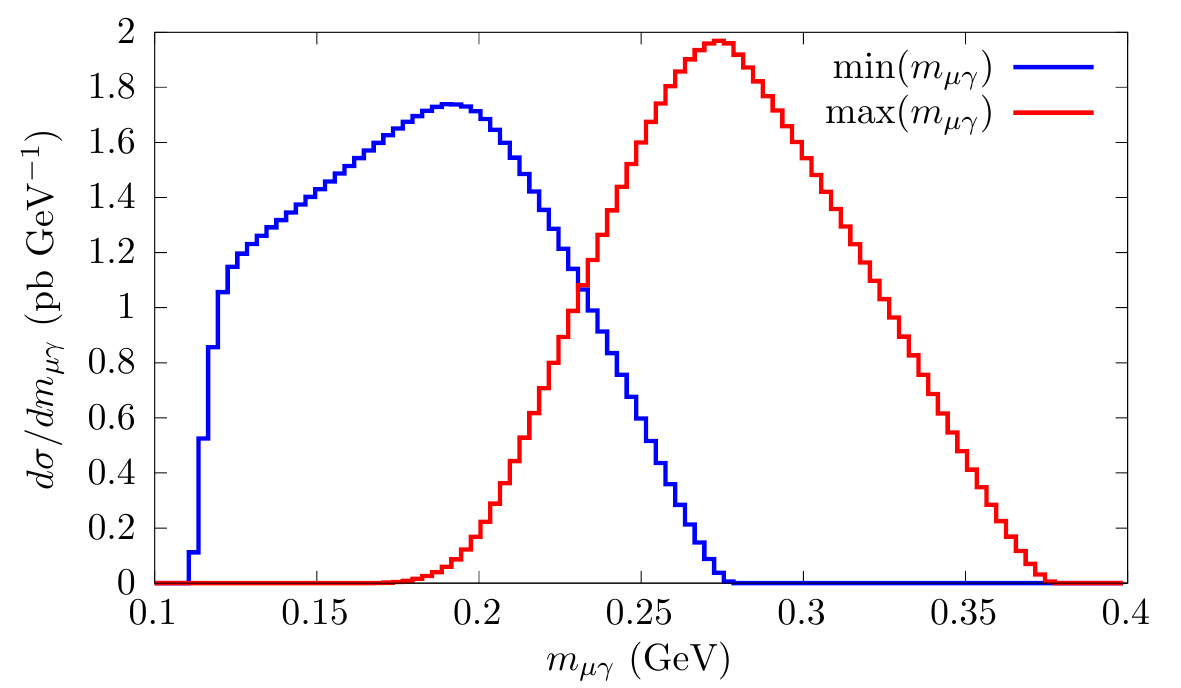}
    \caption{The same as in Fig.~\ref{fig:th_g1g2} but plotted against the minimum (blue curve) and the maximum (red curve) of the invariant mass of the $\mu-\gamma$ system $m_{\mu-\gamma}$.}
    \label{fig:m_mg}
\end{figure}

\begin{figure}[b]
    \centering
    \includegraphics[width=\columnwidth]{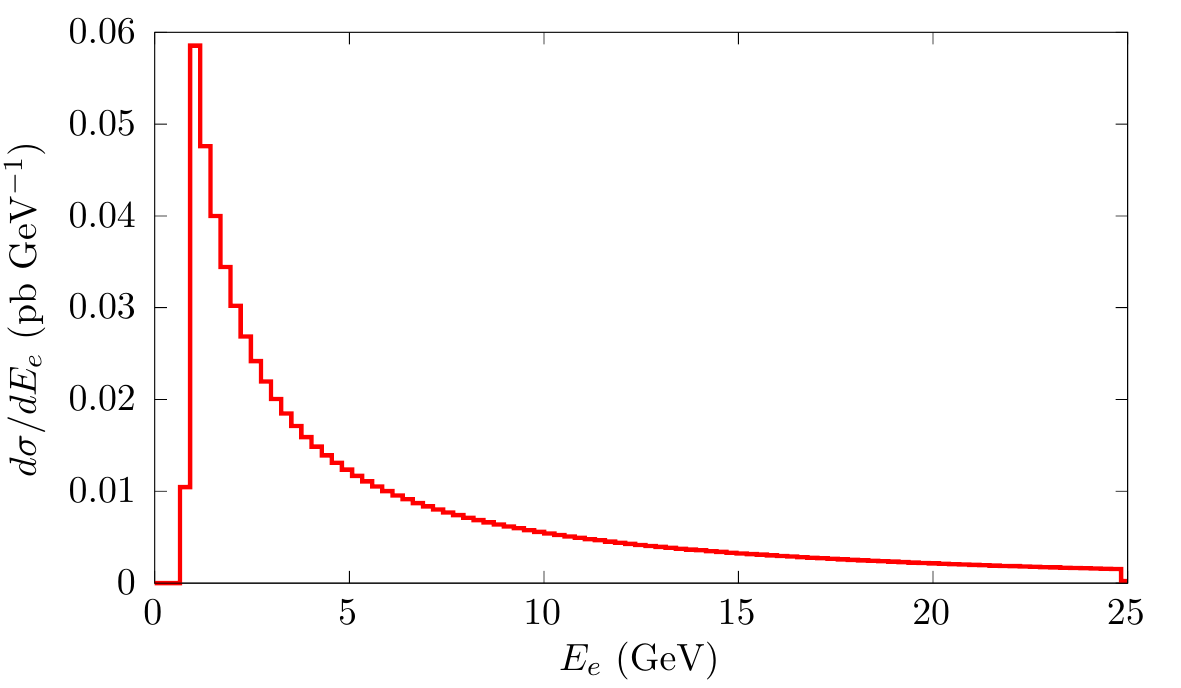}
    \caption{The same as in figure \ref{fig:th_g1g2} but plotted against the outgoing electron energy $E_e$.}
    \label{fig:eenlabj}
\end{figure}

The same line of reasoning is followed for the $\mu-\gamma$ system, with the results shown in Fig.~\ref{fig:m_mg}.
The distribution for the minimum of the invariant mass $m_{\mu\gamma}$ ranges between $0.1$ GeV and $0.28$ GeV, peaking at about $0.19$ GeV with a value for the differential cross section of $1.8$ pb GeV$^{-1}$. The  $\max(m_{\mu\gamma})$ distribution ranges between $0.17$~GeV and $0.38$~GeV with a peak at about $0.27$ GeV, where the distribution reaches $2$ pb GeV$^{-1}$.
Fig.~\ref{fig:eenlabj} shows the distribution for the outgoing electron energy $E_e$. Given the event selection, the range of this observable is between $1$ GeV and $25$ GeV. The bulk of the events have small $E_e$: the peak of the differential cross section is approximately at $1.2$ GeV, where it reaches almost $6\times 10^{-2}$ pb GeV$^{-1}$. This distribution has a shape qualitatively similar to the 
one of the signal process (cfr. Fig.~2 of Ref.~\cite{Asai:2021wzx}), with a 
more pronounced tail after the peak.

\section{Conclusions}\label{sec:conclusion}
In this letter we have discussed the $\pi^0$ production at MUonE 
through the process $\mu e\to \mu e \pi^0$ with the exact tree-level calculation of the matrix element and its implementation 
in the MC generator \textsc{MESMER}. By studying its numerical impact at the differential level on the distributions relevant 
for the measurement of the hadronic contribution to the QED running coupling constant, we can conclude that $\pi^0$ production is a completely negligible reducible background 
in view of a target precision of 10ppm. Considering also that pion pair production is kinematically forbidden for realistic event selections, the present study shows 
that real hadron production in $\mu e$ scattering at the MUonE 
experiment does not affect the measurement of the QED running coupling constant.
The same process $\mu e \to \mu e \pi^0$ can be of interest for New Physics searches at MUonE involving $2 \to 3$ channels, in phase space regions complementary to the one of the $\mu e$ elastic process. For the sake of illustration, we quantified its impact as a background to 
$Z^\prime$ production in a $L_\mu - L_\tau$ gauge model, 
as recently suggested in the literature. Through a MC simulation 
we characterized some relevant distributions involving the photons 
from $\pi^0$ decay, to be considered for a photon veto 
analysis strategy. Further detailed studies could be performed with 
the upgraded MC generator \textsc{MESMER}.

\section{Acknowledgements}
We are sincerely grateful to all our MUonE colleagues for the continuous stimulating 
collaboration. In particular we thank Mauro Chiesa, Guido Montagna and 
Oreste Nicrosini for the valuable discussions and the careful reading of the manuscript.

\appendix

 \bibliographystyle{elsarticle-num} 
 \bibliography{muepi0}

\end{document}